# Long-period surface motion of the multi-patch Mw9.0 Tohoku-Oki earthquake

Panos A. Psimoulis [1,2*], Nicolas Houlié [2,3], Clotaire Michel [4], Michael Meindl [2] and Markus Rothacher [2]

[1] Nottingham Geospatial Institute, Dept. of Civil Engineering, The University of Nottingham, Nottingham, NG7 2TU, UK

[2] Geodesy and Geodynamics Lab., Institute of Geodesy and Photogrammetry, ETH Zurich, Zurich 8093, Switzerland

[3] Seismology and Geodynamics, Institute of Geophysics, ETH Zurich, Zurich 8092, Switzerland

[4] Swiss Seismological Service, ETH Zurich, Zurich 8092, Switzerland

**Abstract**

We show that it is possible to capture the oscillatory ground motion induced by the Tohoku-Oki event for periods ranging from 3 to 100s using Precise Point Positioning (PPP). We find that the ground motions of the sedimentary basins of Japan were large (respectively > 0.15m/s and >0.15m/s$^2$ for velocity and acceleration) even for periods larger than 3s. We compare geodetic observables with a Ground Motion Prediction Equation (GMPE) designed for Japan seismicity and find that the Spectral Acceleration (SA) is well estimated for periods larger than 3s and distances ranging from 100 to 500km. At last, through the analysis of the displacement attenuation plots, we show that the 2011 Tohoku-Oki event is likely composed of multiple rupture patches as suggested before by time-reversal inversions of seismic data.

[*] Corresponding author: Assistant Professor in Geospatial Eng., Tel: +44 (0) 115 951 4287, Fax: +44 (0) 115 951 3881, E-mail: Panagiotis.Psimoulis@nottingham.ac.uk

# 1. Introduction

For solid Earth investigation, high-rate GPS is nowadays an essential tool in the study of transient deformation, seismic source at depth, or accurate determination of strain rate and surface velocity fields. In the field of engineering, high-rate GPS is used to determine the deformation or the motion of structures (bridges, high buildings, etc.) in time (Çelebi and Sanli, 2002; Psimoulis et al., 2008). Koketsu and Miyake (2008) showed that long-period ground motion is also of interest for the increasing number of large civil engineering structures such as bridges and tall buildings since far-distance resonance and short-distance directivity could occur. In the current study we investigate whether GPS is reliable and accurate in the determination of oscillatory ground motions (acceleration, velocity, displacement) for frequencies lower than or equal to 0.33 Hz.

Peak ground accelerations and velocities (PGA, PGV) are routinely computed using seismic data in near real-time (minutes or less) in the 0.5 to 25Hz frequency range (typically) in order to assess the possible consequences of an earthquake shortly after its occurrence. The relationship between magnitude and peak ground motions of an event is distorted by the interaction between seismic waves and the structures (sedimentary basins, fault systems, etc.) sampled during their travel from the source (e.g. Roten et al., 2011 or Denolle et al., 2013). Ground Motion Prediction Equations (GMPEs) predict that the peak motions representative of longer periods, such as the spectral acceleration (SA) at 3s depend proportionally more on the event's magnitudes than PGA or PGV (e.g. Zhao at al., 2006). Nevertheless, few GMPEs for peak motions representative of long-periods are available in the literature due to the limited amount of broadband data (Cauzzi and Faccioli, 2008). For this reason, peak ground displacements (PGDs) are generally not available and proxies like the Displacement Response Spectrum DRS (10s) are used (Cauzzi and Faccioli, 2008). We explore whether GPS can contribute to GMPEs but also if the seismic source can be better characterized using long-period surface oscillatory motions.

As seismic waveforms, GPS time-series provide insights into surface transients observed during seismic wave propagation or during a seismic rupture (Larson, 2001; Bock et al., 2004; Ji et al., 2004; Miyazaki et al., 2004; Blewitt et al., 2006; Kobayashi et al., 2006; Ohta et al., 2006; Emore et al., 2007; Wang et al., 2007; Yokota et al., 2009; Delouis et al., 2010; Feng et al., 2010; Houlié et al., 2011, Bock et al., 2011; Avallone et al., 2011; Ohta et al., 2012; Mitsui and Heki, 2012 Wright et al., 2012; Meng et al., 2013; Guo et al., 2013; Houlié et al., 2014, Psimoulis et al., 2015), tsunami generation for early warning (Blewitt et al., 2006; Sobolev et al., 2007; Blewitt et al., 2009; Crowell et al., 2009; Behrens et al., 2010; Ohta et al., 2012; Li et al., 2013), changes in ionosphere status (Ducic et al., 2003; Mai and Kiang, 2009; Karia et al., 2012; Galvan et al., 2012), volcanic plumes (Houlié et al., 2005; Houlié et al., 2006; Newman, et al., 2012), water vapour delays (Perler et al., 2011; Hurter et al., 2012), or moment tensor (O'Toole et al., 2012; O'Toole et al., 2013). However, the studies listed above did not make use of real-time or near real-time data processing strategies. Precise Point Positioning (PPP) allows to monitor the motion of standalone GPS sites (Blewitt, 2008; Ge et al., 2008; Jokinen et al., 2013). GPS, which is accurate up to a few millimetres for high-frequency oscillations in differential mode (Psimoulis and Stiros, 2008; Psimoulis and Stiros, 2012), is accurate at the centimetre level in PPP mode (Moschas et al., 2014), accuracy which is sufficient to determine the surface displacement (Figure 1) following large events (Mw > 7.0). The PPP GPS time-series have not been used until today as a tool to describe the surface oscillatory motion. In this study we propose to use PPP time-series processed for the GEONET continuous GPS (CGPS) sites to study the ground motions that followed the 2011 Tohoku-Oki event.

The Tohoku-Oki event (origin time $T_0$ at 05:46:23 UTC on March 11, 2011) and the subsequent tsunami stroke Japan with a heavy impact on structures. The rupture area is exceptionally compact for such magnitude (~Mw9.0 and L~500km, to be compared with the 1000 km of rupture of the 2004 Sumatra event with a similar magnitude, Vigny et al., 2005) with most of the slip contained above 30 km depth and a maximum slip on the fault of nearly 50m (Koketsu et al., 2011; Yagi et al., 2011; Suzuki et al., 2011; Yue and Lay, 2011). Because of the poor coverage near the epicentre, the slip distribution near the surface is not well constrained. Other studies (Maercklin et al., 2012; Roten et al., 2012) confirmed that the long-period (10-20s) maximum slip was up to 50m but also suggested that the Tohoku-Oki event was indeed composed of multiple ruptures of smaller magnitudes (<Mw8.8).

**2. Data**

**2.1 GPS data**

The Mw9.0 earthquake off the Pacific coast of Tohoku-Oki on March 11, 2011 was fully recorded by the GEONET network (Sagiya, 2004), operated by the Geospatial Information Authority of Japan (GSI). The GEONET network is composed of over 1200 continuously observing GPS receivers (with an average spacing of 20km). We processed 1Hz GPS records from 847 GEONET stations of 15-hour duration, fully covering the earthquake period. The data was analysed with the Bernese GPS Software (Dach et al., 2007). The data were post-processed in a PPP mode using state-of-the-art models and a-priori information of highest quality from the Center for Orbit Determination in Europe (Dach et al. 2009, Bock et al., 2009). As an example for the results obtained, Figure 2 shows the displacement field at 300 seconds after the origin time ($T_0$). The displacement time series in North, East, and vertical components were established separately for each station. The GPS displacement time series show an *a-posteriori* formal accuracy of about 1cm in the horizontal and 2cm in the vertical component, respectively (Psimoulis et al., 2015).

**2.2 Strong motion data**

The largest strong-motion networks of Japan called K-NET (Kyoshin network) and KiK-net (Kiban-Kyoshin network) consist of 1034 and 660 stations, respectively (Aoi et al., 2011). The main difference of the two networks is that the K-NET stations are located mainly on thick sedimentary sites, while the KiK-net stations are deployed on rock or thin sedimentary sites, as it is a sub-net of the Hi-net (High-sensitivity seismograph network) primarily designed for highly sensitive seismic observations. Consequently, the K-NET and KiK-net triggering thresholds are different (2 cm/s$^2$ and 0.2 cm/s$^2$, Aoi et al., 2004). Furthermore, the K-NET stations are installed on the ground surface, while each KiK-net site hosts two instruments, one installed at the surface and the second at the bottom of a borehole of 100-200m depth.

The Mw9.0 earthquake of Tohoku-Oki 2011 was well recorded by the two strong-motion networks. Raw data of strong-motion records from 700 K-NET sites and 525 KiK-net sites were available. The records of three components, corresponding to the North, East and vertical directions, were sampled at 100Hz and up to 300s duration. The acceleration of each site was derived by correcting the raw data for the gain, converting the corresponding record time from UTC to GPS time and then by applying the time correction for 15 leap seconds. In order to be consistent with the recorded motion of the GPS and K-NET sites, only KiK-net surface records were investigated in this study.

**3. Results**

**3.1 Consistency of GPS and strong-motion data**

The evaluation of the consistency of GPS and strong-motion records at high frequencies must be completed for sites located very closely. In previous studies, the GPS and strong-motion sites with an inter-distance of less than 1km were assumed as collocated (Melgar et al., 2013). In our study, we used an even more conservative approach by using only the very closely located GPS and strong-motion sites, with an inter-distance less than 100m. In total, we identified 29 very closely located sites of the GPS and strong-motion networks (Figure 3), comprising 21 K-NET and 8 KiK-net sites (Table 1).

For these sites we assume that the GPS and strong-motion records are not shifted by a significant time difference (<1s), keeping in mind that the sampling rate (1Hz) of the GPS network is low compared to the sampling rate of the strong-motion network (100Hz).

The evaluation of the consistency of the GPS and the strong-motion sensors was examined by comparing the spectra of the two records and assessing them in their common frequency range (0-0.5Hz). The usual technique of integration of strong-motion records for the computation of velocities and displacements was not followed due to the significant accumulated error (Stiros, 2008), the large duration of the strong-motion records (up to 300s) and the significant coseismic displacement (up to 2-3m), which would make the estimation of the long-period displacement time series unreliable (Boore, 2003).

Thus, instead of integrating the strong-motion records, the GPS displacement series were differentiated twice resulting in acceleration time series. The derived accelerations are affected also by errors due to the differentiation, which, however, do not accumulate. These differentiation errors affect mainly the short-period acceleration (2-3s), due to the high short-period noise of GPS measurements relatively to the short time step of the differentiation. Furthermore, the error of the derived accelerations is expected to be higher for the vertical component due to the higher noise level of the corresponding PPP results.

In Figure 4, we present the spectra of the GPS and the K-NET acceleration time series of the East-West (EW), North-South (NS), and Up-Down (UD) components for two representative closely located sites, one close (179km; Figure 4a-c) and one far (771km; Figure 4d-f) from the epicentre. In addition, the spectra of the GPS acceleration time series, corresponding to the time interval before the seismic event, are shown in Figure 4 as black lines. These spectra provide information on the basic noise of the acceleration time series derived from GPS. For periods above 3-4 seconds, the GPS noise is lower than the seismic signal for the horizontal component. This supports the consistency of GPS and K-NET spectra (Figure 4a-c) mainly for periods above 3 seconds, while KiK-net strong-motion spectra prove to be slightly less consistent (Figure 4d-f) for the periods below 5 seconds. This is the case in general for sites far from the epicentre (>600 km) where the seismic signal is weak relative to the PPP noise and the S/N decreases, especially for short-periods (<5s) and for the vertical component (period <10s). However, such phenomena do not affect the current study, as it is focusing on the use of the horizontal components and long-periodic oscillations (>3 s), which usually are not included in the ground motion parameters computation.

## 3.2 Computation of MGA, MGV and MGD

The Maximum Ground Acceleration (MGA), Velocity (MGV), and Displacement (MGD) were computed from GPS and strong-motion records for short- (3-10s) and long-period (10-100s) domains. The strong-motion records were integrated once and twice into velocities and displacements and the GPS records were differentiated once and twice to obtain velocities and accelerations. The derived strong-motion and GPS time series were filtered for a short- (3-10s) and a long-period (10-100s) band. Figure 5 shows the acceleration time series of the GPS and strong-motion instruments that are closely located for the two period bands. For each type of sensor, we compute the absolute maxima (corresponding to MGA, MGV, and MGD) of acceleration, velocity, and displacement time series of every site of each network. Figures 6a, 6b and 6c show the MGA, MGV and MGD maps for the two period bands. As expected, by comparing the GPS and the strong-motion networks, GPS seems to be more sensitive to long-period ground displacements and the strong-motion sensors to short-period accelerations.

The noise of the GPS acceleration, velocity, and displacement time series was estimated by analysing the time interval before the seismic event, where all the sites are assumed not to be affected by motions. Following the methodology of MGA, MGV and MGD, the GPS noise displacement time series were differentiated once and twice and subsequently filtered for the two period bands (3-10s and 10-100s). The maximum value of each resulting time series expresses the maximum noise level for each GPS site and for each quantity (acceleration, velocity, displacement). In Table 2 the mean average and the standard deviation of the estimated maximum noise values for the horizontal component and for the two examined period bands are listed. All the estimated MGA, MGV, and MGD values of the GPS sites are above their respective uncertainty maxima, with the latter computed as the mean maximum error+3σ zone (2.0cm/s$^2$, 1.5cm/s, 1.0cm, Figure 7). The level of the estimated maximum noise is smaller than that in other studies (e.g. Wright et al., 2012), due to the limitation of the noise level by filtering the GPS time series for specific period bands.

The qualitative examination of Figure 6 shows that strong-motion and GPS provide a consistent estimation of MGA, MGV and MGD for the two examined period domains. Some local differences for the short-period band might reflect weaknesses of GPS in estimating acceleration of low periods (<4-5s) or local effects, without disturbing though the overall compatibility of the MGA maps.

At long-periods, the maximum displacement map (Figure 6c, right) is smooth and reflects the distance to the largest slip patch on the fault: the maximum values occur close to the epicentre in the Sendai region with large values along the coast to the South. However, in the 3-10s band (Figure 6b, c, left), the map is more complex and highlights large sedimentary basins. In addition to the Sendai basin, close to the epicentre, the largest values are found in the Kanto plain, known to exhibit resonance frequencies in this band (Yamanaka et al., 1989). On the western coast, large values can also be found at the Sakata and Niigata basins, in relation with the large amplification in these frequency band as shown by Mamula et al. (1984).

In Figure 8, we show the differences of the estimated GPS and strong-motion MGA, MGV and MGD for the very closely located sites, normalised to the maximum acceleration, velocity and displacement of the corresponding site. The derived normalised differences show the good agreement between the solutions of GPS and the strong-motion sensors (relative errors mostly below 10%) with the exception the short-period (3-10sec) MGA and MGD differences. The latter are observed mainly for large distances from the epicentre and they are caused by the relatively small oscillatory amplitude of the corresponding accelerations and displacements (up to $0.15 m/s^2$ and $0.10 m$, respectively). This can be seen more clearly from the absolute differences of the estimated MGA, MGV and MGD between the GPS and strong-motion sensors in Figure A1, which reveals the small values ($<0.03 cm/s^2$, $<0.03 cm/s$ and $0.05 cm$, respectively) for the sites located mainly far from the epicentre. The relatively high differences in the displacement (>3cm) of some sites far from the epicentre (K-NET GIF006 - GPS 0280, K-NET ISK015 - GPS 0972, K-NET HKD043 - GPS 0502 and K-NET HKD046 - GPS 0504) may be attributed to local effects.

Since we find a general agreement of the GPS records with the seismic data in space and frequency, we compute the spectral acceleration (SA) at 3s for a 5% damping (Figure 9). In general the long-period signals are not of concern for the safety of typical building or small objects at the surface. However, a large object such as a bridge, a skyscraper or a deep basin might resonate at long-periods, particularly in the 3-10s range (Anderson et al., 1985; Singh et al. 1988; Lermo et al., 1994; Chávez-García and Bard, 1994). Even more rigid structures might be affected by these long-periods due to site effects, soil-structure interaction and liquefaction (Margaris et al., 2010; Assimaki et al., 2012). In order to have a baseline in the current study, the attenuation of ground motion with distance (spectral acceleration and acceleration) is compared to the GMPE proposed for Japan by Zhao et al. (2006). Please note that we use the distance from the epicentre (as computed by USGS) whereas Zhao et al. (2006) assume the shortest distance between a rupture model and the station. We find that SA(3s) estimates from GPS and strong-motion data are compatible with maximum values predicted by the GMPE (Figure 9).

Figure 10 shows the strong-motion and GPS maximum accelerations, velocities and displacements of the two period bands (3-10s and 10-100s) with respect to the distance to the epicentre. The comparison of MGA with GMPE predictions is less relevant as the long-period signals were not used to build the GMPE. For large distances, several values of MGA exceed the estimated PGA predicted by the GMPE. We remind that this discrepancy is however related to small amplitude motions and more interestingly to the response of basins located at distance > 200km. Basins response is however not sufficient to explain the evolution of MGA, MGV and MGD with distance.

Like the accelerations, the maximum displacement estimates depend on the period range. We know that MGD for the short-periods are ~10 times smaller than for long-periods (at ~1m), and we might expect that PGD follows the same trend, so does the PGA. In the case of a single rupture patch, we expect a decay of the form 1/distance in the displacement/distance diagrams. In Figure 10f, displaying long-period oscillatory displacements, the displacement vs. distance curve does not constantly decrease, but also increases with a second peak at a distance of about 250 km from the epicentre. For better visibility, the averaged displacement for each distance is shown using a blue line. Such a pattern for long-periods (10-100s) suggests that the ground motion results from the rupture of two distinct fault segments. The averaged velocity and acceleration is also computed (blue line, Figure 10d, e) but the feature of increment of the two variables is not so pronounced mainly due to the weak velocity and acceleration signal in long-periods.

Regarding the oscillatory acceleration and velocity, they appear to be constant for the short-period (3-10s) and distance up to 250-300km from the epicentre (Figure 10), and then decrease with distance linearly (in a log-log space). This pattern suggests that the strong short-period oscillatory component is generated in an extended rupture area, which seems to cover the location of the two different rupture patches, as it was revealed by the long-period displacement. The plateau cannot be seen clearly in the short-period oscillatory displacements due to the relatively small signal of the corresponding displacements.

Stewart et al. (2013) extensively compared GMPEs to the observed ground motion of the Tohoku-Oki event and could not detect the stability of MGA/MGV with increasing distance. Indeed, whereas high-frequency ground motion (intensity measures such as PGA or PGV) is generated by small-scale asperities all along the fault, low-frequency parameters of the ground motion such as displacement is caused by large-scale asperities that may be more localised on the fault as shown by source inversions. These phenomena impact the choice of an optimal distance measure: whereas distance to rupture is the most relevant for high frequencies as used by Stewart et al. (2013), distance to major asperities is better for low frequencies. The possibility of the presence of two sources is critical for correct ground motion estimation. Indeed, as the two sources would be shifted relatively to each other, expected ground motion would be larger at larger distances from the first epicentre.

## 4. Conclusions and outlook

The first conclusion of our work is that both GPS displacement histories and integrated accelerograms are in agreement for the frequency range they have in common (3 to 100s). We find that long-period oscillatory ground motions are accurately determined when based on GPS data. The double integration of accelerograms does not impact the precise resolution of the derived displacement data for periods less than 100s for a large earthquake such as the Tohoku-Oki event.

In future, the PPP GPS data will be processed in near real-time or even real-time (Temporal Point Positioning; Li et al., 2013), allowing the computation of maximum ground motion to supplement other seismic monitoring systems (Liu et al., 2014; Tu, 2014). Considering that the ground acceleration is expected to be more than 5 times higher at a period of 1s than the signal at a period of 10s, the reliability of the GPS for higher sampling rates (>1Hz) will mostly depend on our capability to maintain the noise of the time series at a level close to cm as suggested by Moschas et al. (2014).

Cauzzi and Faccioli (2008) demonstrated the importance of accurate GMPEs at long-periods and the limited amount of available seismic data at these periods. We propose that the future generations of GMPEs should include more long-period data such as GPS displacement time-series, in order to better estimate effects of large magnitude (Mw>7) events.

Furthermore, we highlight the response of major sedimentary basins of Japan to the shake that followed the Mw9.0 Tohoku-Oki 2011 earthquake, for periods ranging from 3 to 10 seconds. The increase in oscillatory displacement at ~250 km distance from the epicentre can be interpreted as a second source of displacement along the fault generated by a second asperity as proposed by the source models of Koketsu et al. (2011), Suzuki et al. (2011) and Maercklin et al. (2012). Long-period maximum oscillatory displacement can therefore be used as a constraint for source inversions, providing additional constraints on spatial pattern of rupture at depth.

**Acknowledgments**

This study has been supported by Swiss National Fund (SNF) grants in the framework of the project "High-rate GNSS for Seismology" (number 200021_130061).

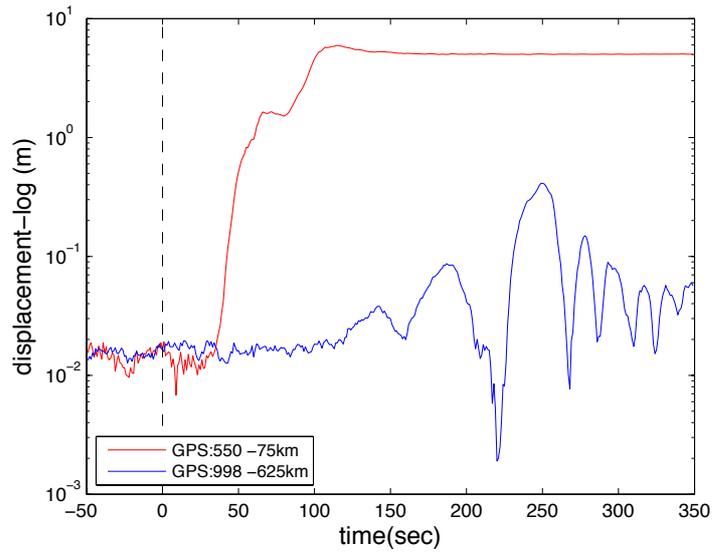

**Figure 1**: The EW displacement time series for the GPS site closest to the epicentre (GPS 0550; ~75km away) and for a remote site (GPS 0998; ~650km); time origin corresponds to the earthquake occurrence. The time difference between the two responses reflects the delay of the arrival of the seismic signal at the station far from the epicentre.

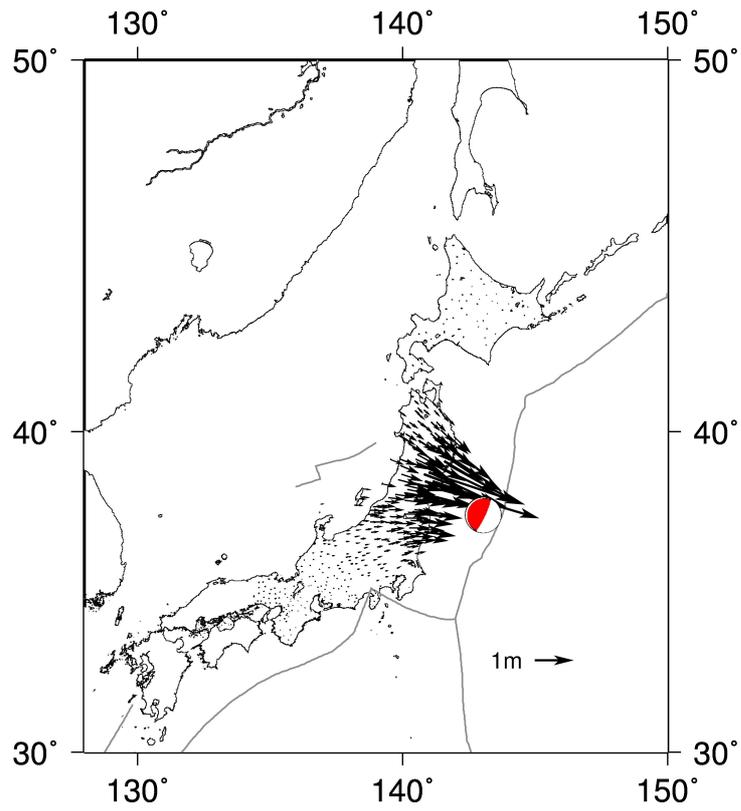

**Figure 2**: Precise Point Positioning (PPP) coseismic horizontal displacement field for the Tohoku-Oki event measured by the GEONET GPS network at t = $T_0$ +300s.

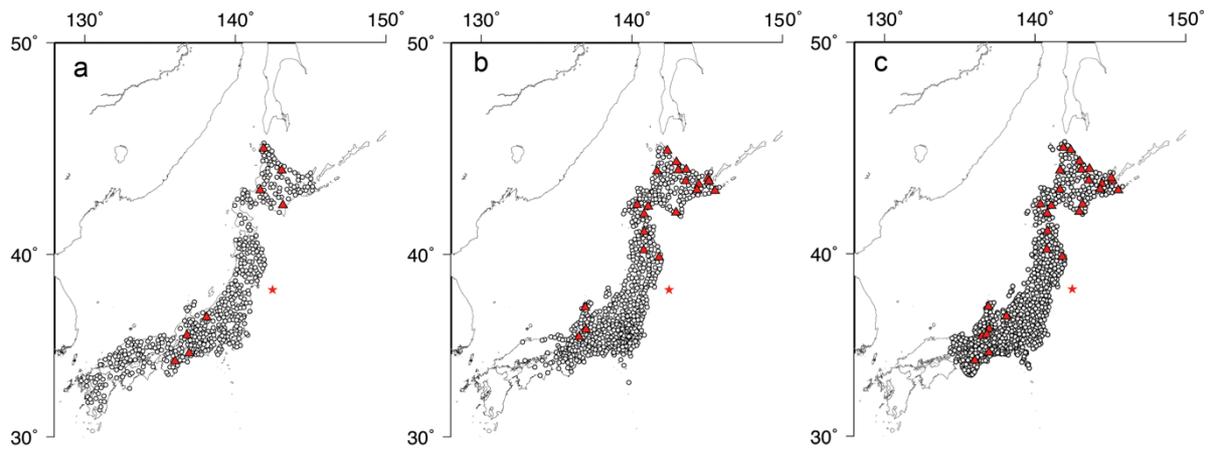

**Figure 3**: KiK-net (a), K-NET (b) and GEONET GPS (c) sites used in this study. The very closely located sites of GPS and strong-motion networks are indicated with red triangles. The red star indicates the epicentre of the Tohoku-Oki earthquake in 2011.

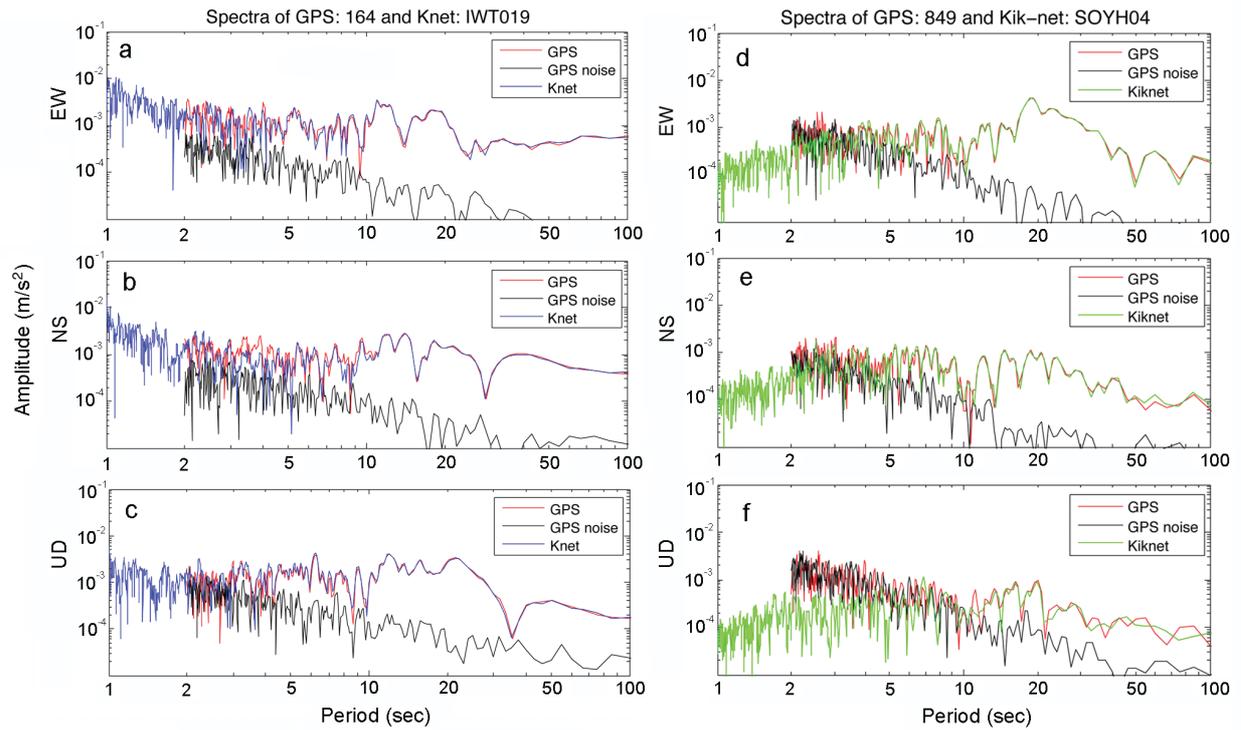

**Figure 4**: Spectra of derived acceleration time series using K-NET, KiK-net and GPS data for two sites: (a-c) close (179km) and (d-f) far (771km) from the epicentre. The solid black line corresponds to the spectra of the GPS acceleration time series before the occurrence of the earthquake indicating the noise level.

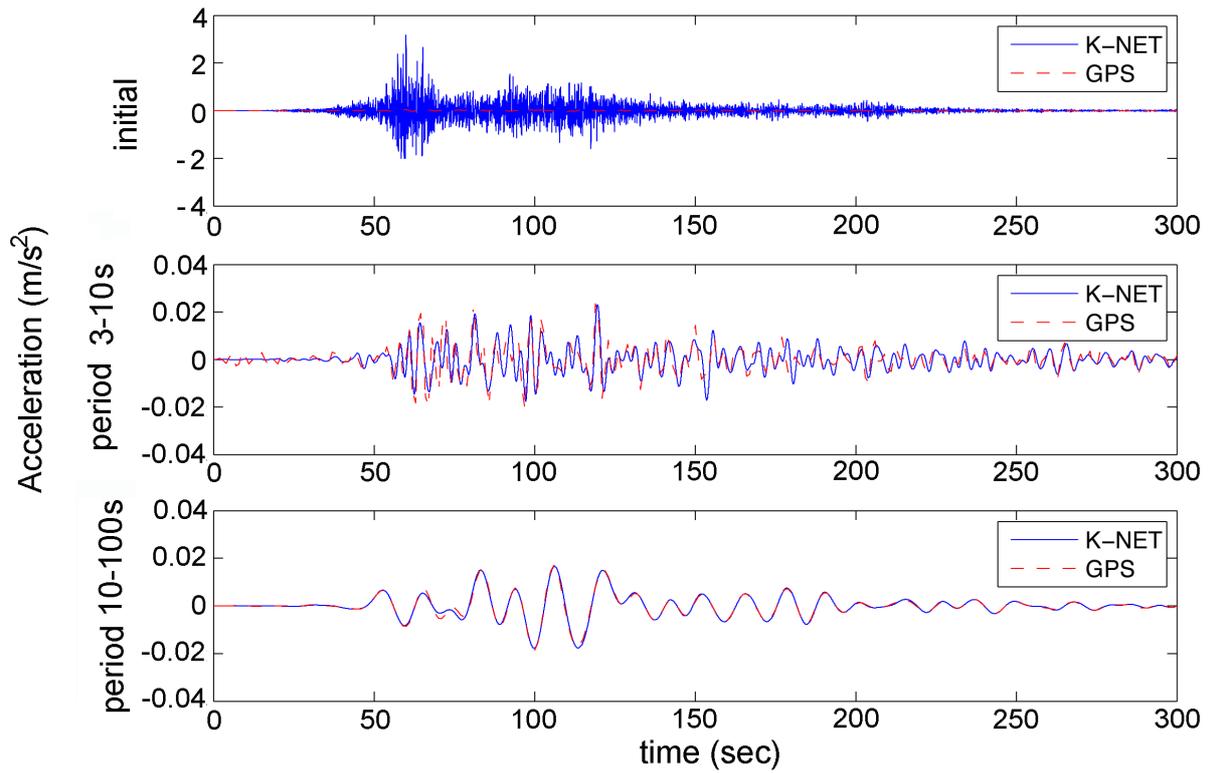

**Figure 5**: The initial acceleration time series of the very closely located sites of K-NET IWT019 (blue solid line) and GPS 0164 (dashed red line) and the derived time series after filtering for the 3-10s and 10-100s period bands. The time corresponds to seconds since the earthquake occurrence.

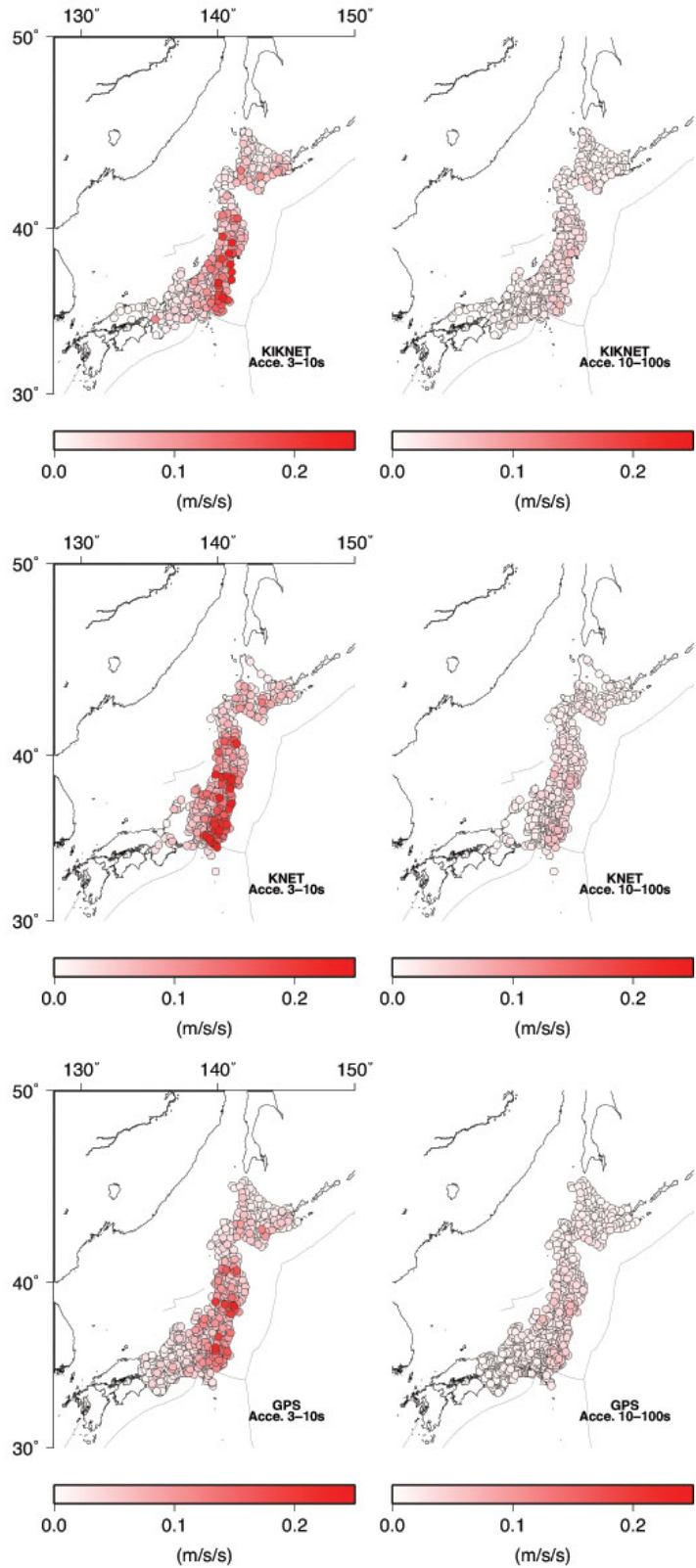

**Figure 6a**: Maximum ground acceleration maps for KiK-net (top), K-NET (middle), and GPS (bottom) for the two period bands (3-10s and 10-100s).

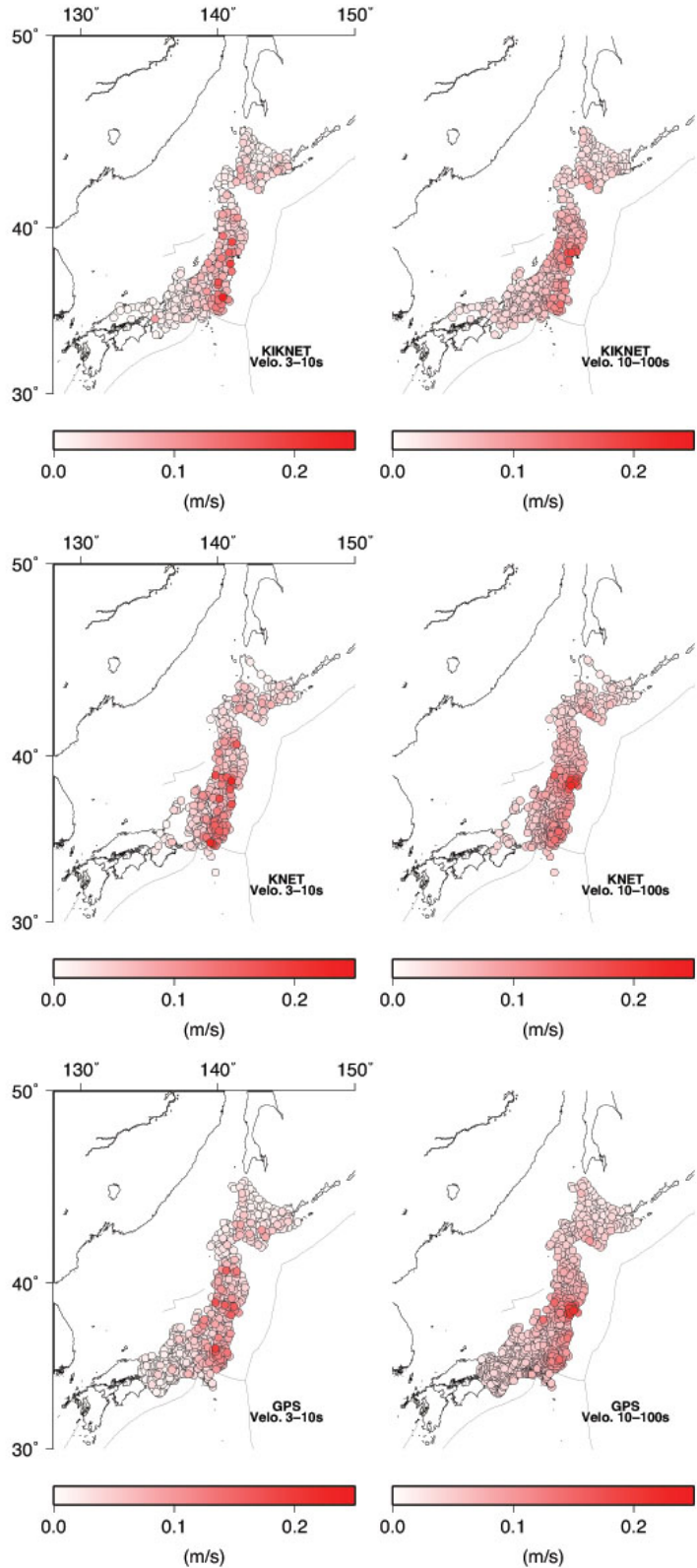

**Figure 6b**: Maximum ground velocity maps for KiK-net (top), K-NET (middle), and GPS (bottom) for the two period bands (3-10s and 10-100s).

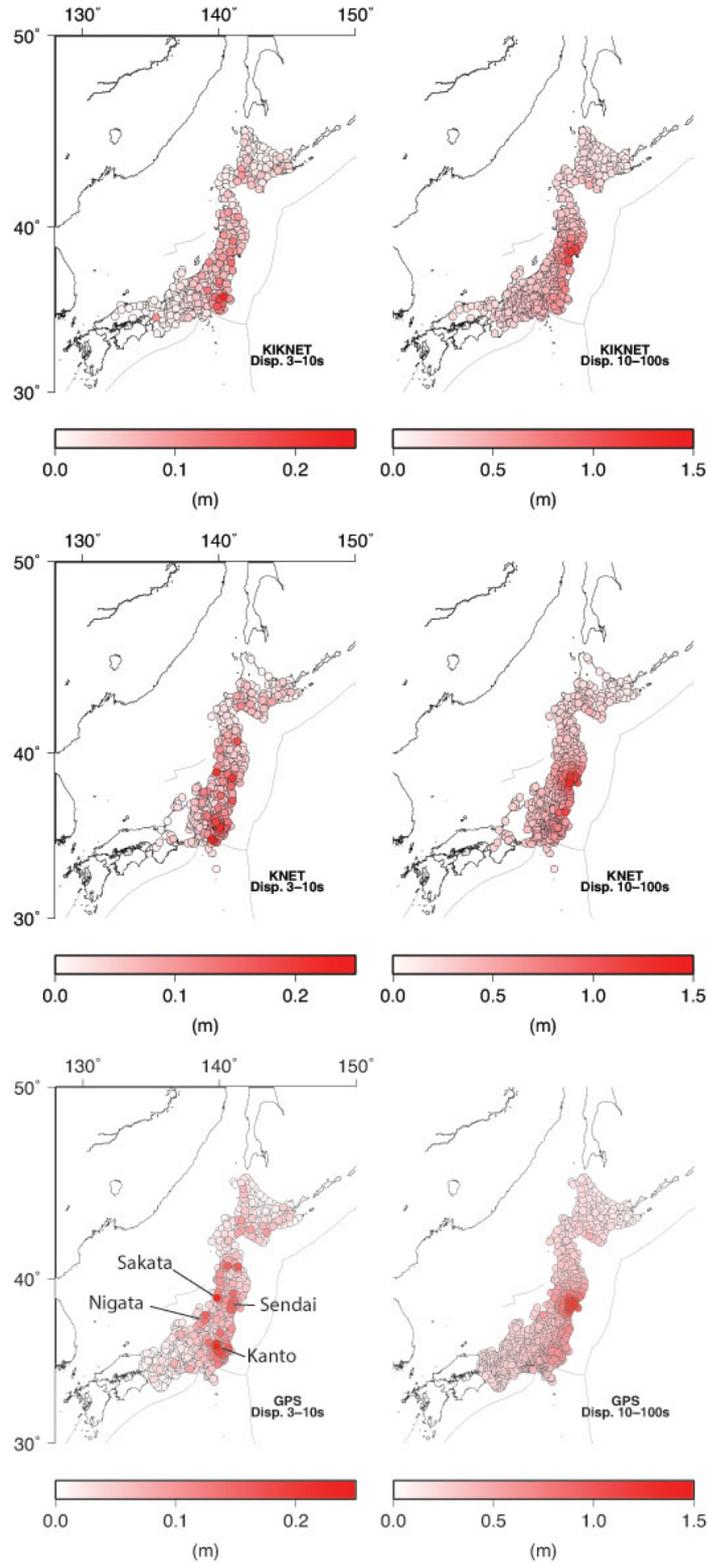

**Figure 6c**: Maximum ground displacement maps for KiK-net (top), K-NET (middle), and GPS (bottom) for the two period bands (3-10s and 10-100s). Note the different scale of the 10-100s band.

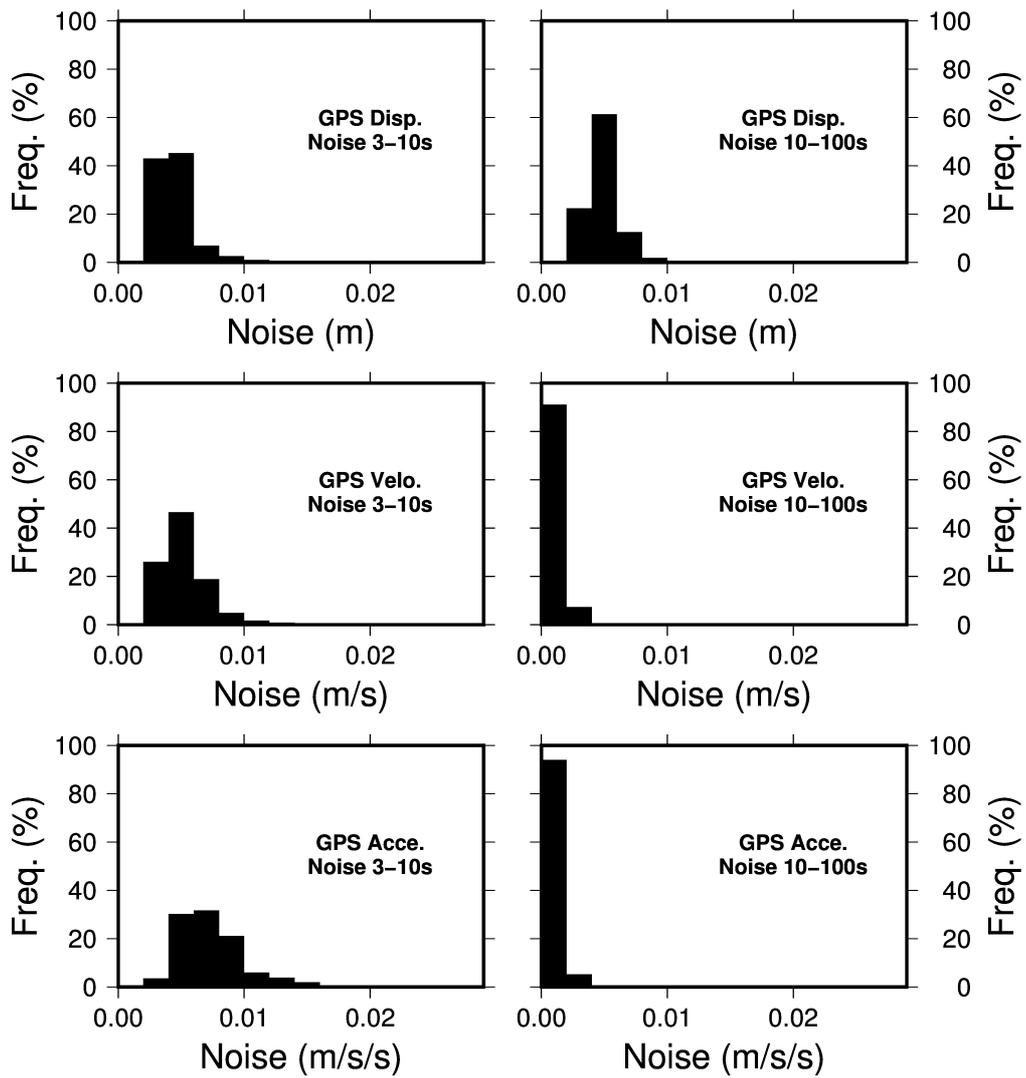

Figure 7: Maximum level of noise derived from the GPS acceleration, velocity and displacement records.

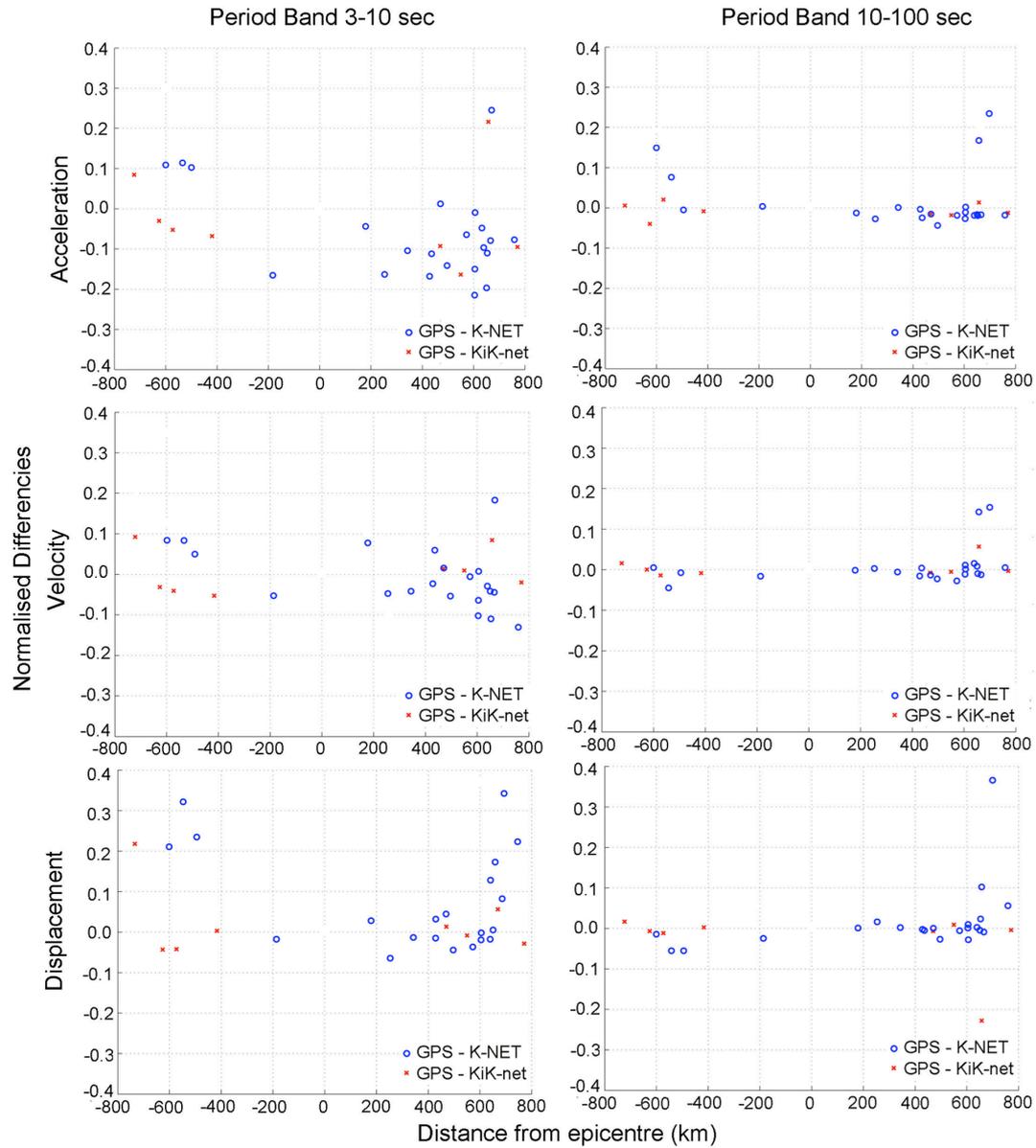

**Figure 8**: Normalised differences of GPS and strong-motion MGA (top), MGV (middle) and MGD (bottom) at the very closely located sites. The positive and negative distances correspond to the position of the sites northward and southward from the epicentre.

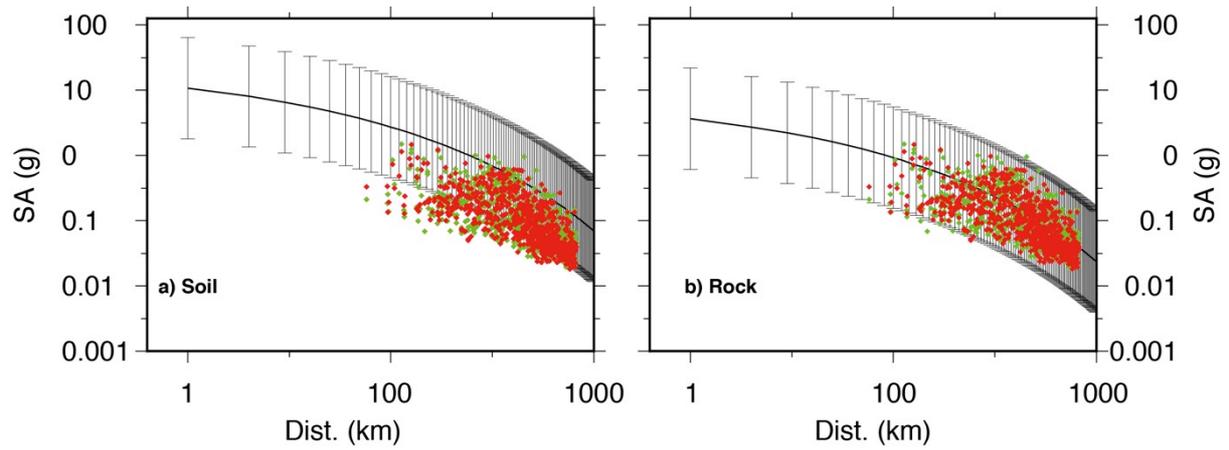

**Figure 9**: Spectral acceleration (SA) at 3s from GMPE (black line) of soft soil (a) and bedrock (b) conditions (Zhao et al., 2006) compared to the SA computed from GPS (red dots) and to the SA computed from the strong-motion records (green dots). The error bars indicate the 1σ uncertainty zone of the GMPE.

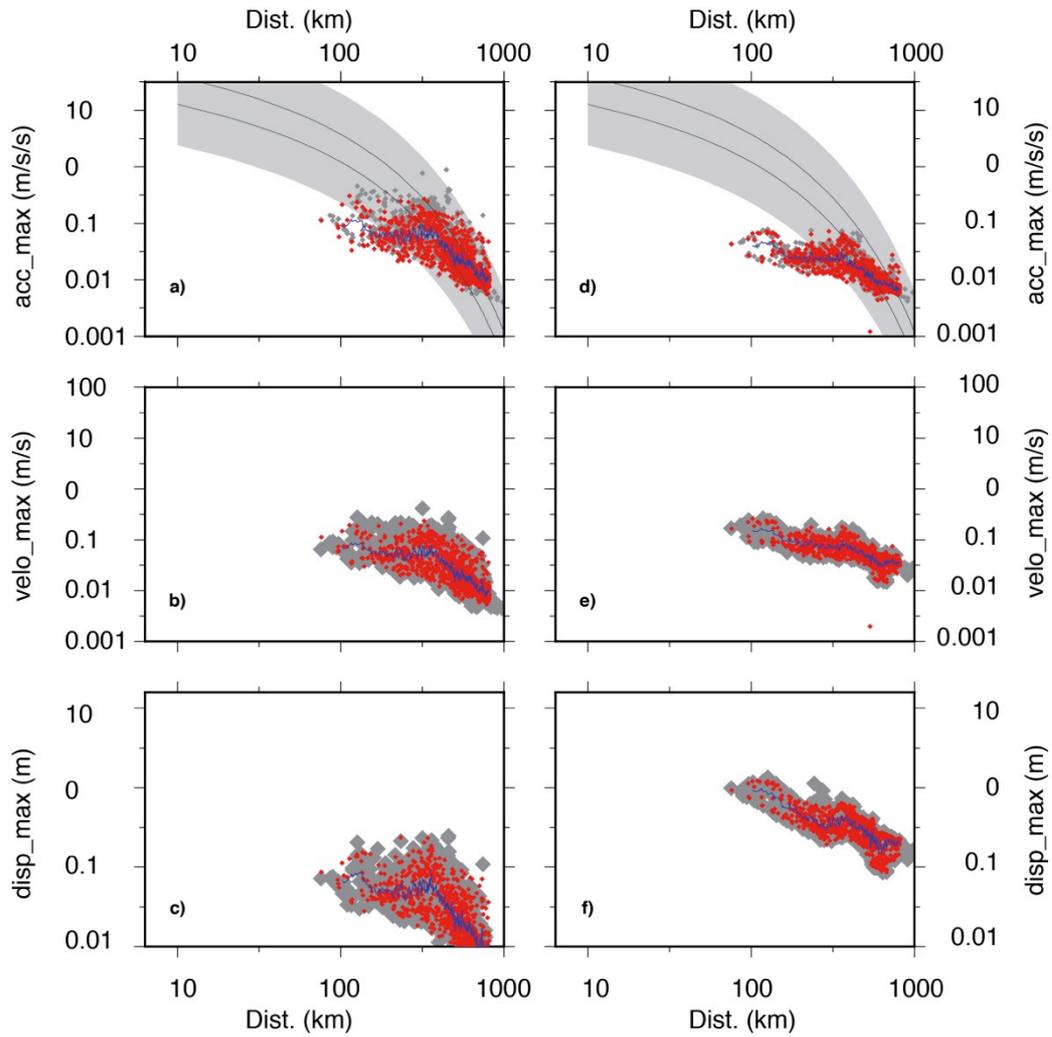

**Figure 10**: GPS (red) and strong-motion (grey) MGA, MGV and MGD for the 3-10s (left) and 10-100s (right) period bands. The blue curve shows the average GPS MGA/MGV/MGD values.

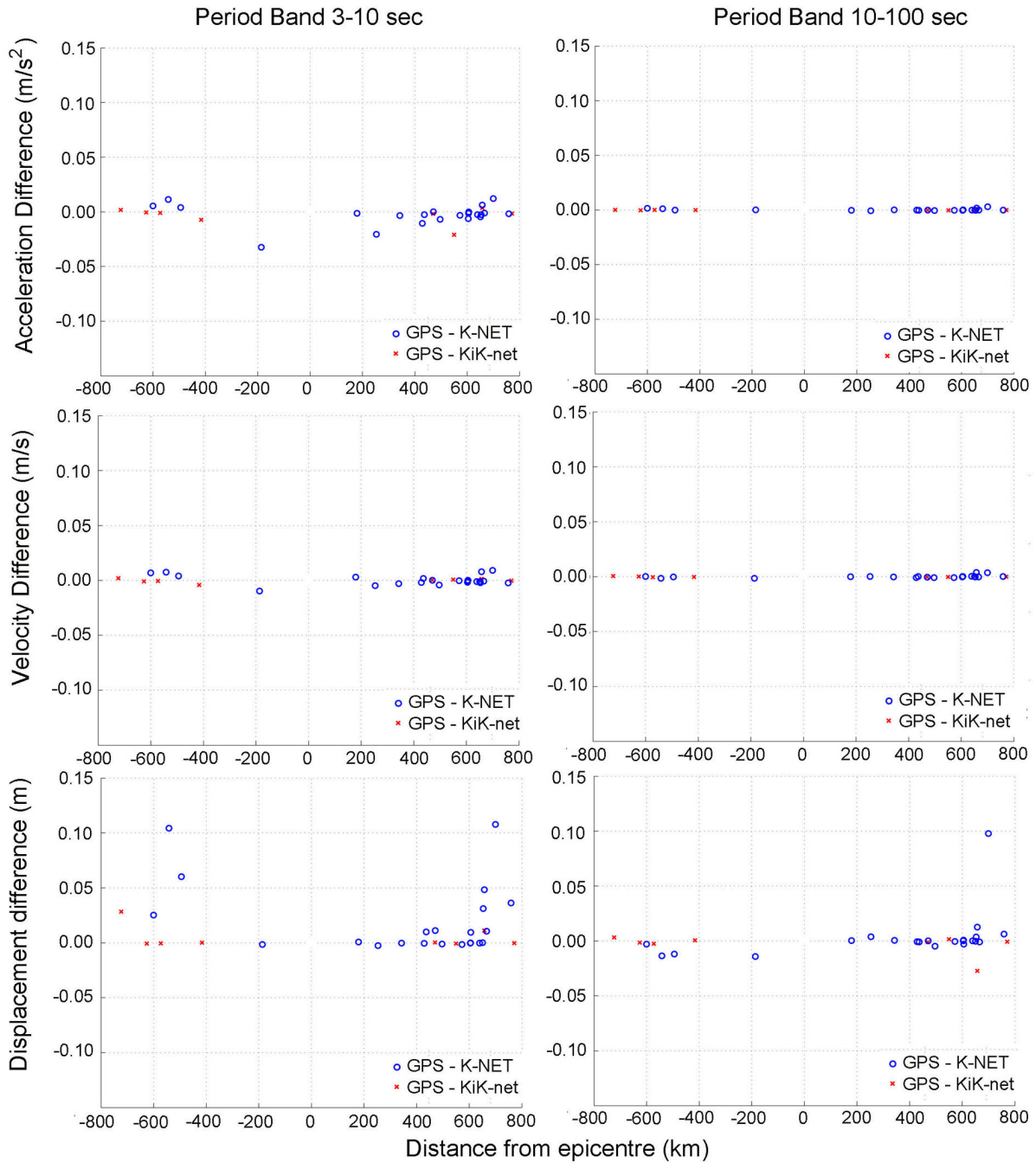

**Figure A1**: The differences of the GPS and strong-motion MGA (top), MGV (middle) and MGD (bottom) values for each very closely located sites versus the distance from the epicentre. The positive and negative distances correspond to the position of the sites northward and southward from the epicentre.

**Table 1**: The closely located GPS and strong-motion sites of K-NET and KiK-net with their location (latitude, longitude) and distance from the epicentre.

| GPS | K-NET | KiK-net | Latitude (degrees) | Longitude (degrees) | Sensor distance (m) | Epicentre distance (km) |
|---|---|---|---|---|---|---|
| 0005 | HKD080 | - | 43.50786 | 144.44901 | 45 | 604 |
| 0041 | FKS011 | - | 37.09071 | 140.90252 | 97 | 186 |
| 0115 | HKD066 | - | 43.66182 | 145.13143 | 20 | 639 |
| 0122 | HKD083 | - | 43.23294 | 144.32503 | 20 | 572 |
| 0140 | HKD151 | - | 42.49436 | 140.35420 | 25 | 496 |
| 0144 | HKD110 | - | 42.13094 | 142.93539 | 39 | 428 |
| 0147 | HKD155 | - | 42.04297 | 140.80554 | 99 | 436 |
| 0164 | IWT019 | - | 39.84918 | 141.80385 | 39 | 179 |
| 0183 | AKT006 | - | 40.21544 | 140.78733 | 26 | 253 |
| 0280 | GIF006 | - | 36.03305 | 136.95276 | 28 | 541 |
| 0285 | GIF012 | - | 35.63615 | 136.48877 | 64 | 600 |
| 0502 | HKD043 | - | 44.58204 | 142.96446 | 6 | 700 |
| 0503 | HKD048 | - | 44.21972 | 143.61567 | 24 | 666 |
| 0504 | - | ABSH04 | 44.19194 | 143.07682 | 12 | 657 |
| 0511 | HKD056 | - | 43.67050 | 143.57815 | 25 | 605 |
| 0519 | HKD072 | - | 43.19517 | 145.52050 | 52 | 605 |
| 0535 | AOM027 | - | 41.14555 | 140.82198 | 29 | 343 |
| 0779 | HKD006 | - | 45.12678 | 142.35184 | 21 | 758 |
| 0783 | HKD020 | - | 44.14876 | 141.66460 | 37 | 652 |
| 0792 | HKD131 | - | 42.42055 | 141.08106 | 67 | 471 |
| 0793 | - | TKCH08 | 42.48641 | 143.15197 | 10 | 470 |
| 0849 | - | SOYH04 | 45.23011 | 141.88178 | 95 | 771 |
| 0864 | HKD065 | - | 43.79403 | 145.05686 | 39 | 650 |
| 0877 | - | IKRH02 | 43.22090 | 141.65202 | 60 | 550 |
| 0972 | ISK015 | - | 37.22644 | 136.90877 | 9 | 494 |
| 0983 | - | NGNH28 | 36.70665 | 138.09673 | 99 | 416 |
| 0991 | - | GIFH23 | 35.72349 | 136.78462 | 7 | 572 |
| 0998 | - | AICH21 | 34.74005 | 136.93848 | 6 | 625 |
| 1009 | - | NARH03 | 34.29293 | 136.00264 | 63 | 723 |

**Table 2**: Mean average and standard deviation of the estimated maximum noise values of the displacements, velocities and accelerations derived from the PPP time series for the two examined period bands.

|  | period 3-10s | | period 10-100s | |
|---|---|---|---|---|
|  | mean average | standard deviation | mean average | standard deviation |
| Displacement (mm) | 4.5 | 2.3 | 5.1 | 2.1 |
| Velocity (mm/s) | 5.5 | 2.7 | 1.5 | 0.9 |
| Acceleration (mm/s$^2$) | 7.6 | 4.0 | 1.0 | 0.7 |